\documentclass[twocolumn,aps,prd,nofootinbib]
{revtex4-1}

\def\K{K{\"a}hler}
\newcommand{\be}{\begin{eqnarray}}
\newcommand{\ee}{\end{eqnarray}}
\def\ba{\begin{array}}
\def\ea{\end{array}}

\def\ba{\begin{eqnarray}}
\def\ea{\end{eqnarray}}

\begin{document}

\title{ \Large\bf Chaotic inflation in supergravity and cosmic string production}

%\title{ \Large\bf Cosmic string production in SUGRA inflation}

\author{Andrei Linde}

\affiliation{Department of Physics, Stanford University, Stanford, CA 94305, USA}

\begin{abstract}
We describe a simple model of chaotic inflation in supergravity where one can easily tune not only the basic cosmological parameters such as $n_{s}$, $r$, and $f_{{NL}}$, but also a possible (subdominant) cosmic string contribution to the CMB anisotropy.
\end{abstract}

\maketitle

\section{Introduction} \label{intro}

Recent observational data impose very stringent constraints on the possible (subdominant) role of cosmic strings in generation of CMB anisotropy. But even thought cosmic strings cannot replace inflation as the main source of perturbations of metric \cite{Pert}, they may induce additional features in CMB maps and serve as a source of non-Gaussian perturbations in the early universe, see e.g. \cite{Ringeval:2010ca,Hindmarsh:2011qj,Shlaer:2012rj,Ringeval:2012tk}.

The standard textbook theory of cosmic string production is related to the theory of high temperature phase transitions in the early universe. But the temperature of the universe after post-inflationary reheating may be too small to produce heavy strings. Recent resurgence of interest to cosmic strings  was based in part on realization that cosmic strings can be produced after inflation in some versions of string theory \cite{Kachru:2003sx,Copeland:2003bj}. Moreover, cosmic strings are naturally produced at the waterfall stage at the end of hybrid inflation \cite{Linde:1991km}. In the context of supergravity, this can be achieved, for example, in the F-term or D-term hybrid inflation \cite{Copeland:1994vg,Binetruy:1996xj}. Unfortunately, in basic versions of these models the contribution of cosmic strings to the CMB anisotropy is unacceptably large, so they require certain modifications for reducing or even eliminating cosmic string effects, see e.g. \cite{Lazarides:2000ck,Haack:2008yb,Buchmuller:2012ex} and references therein. 

All of these models of cosmic string production {\it after inflation} are vulnerable with respect to the constraints that can be placed on the cosmic string tension by pulsar timing limits on the gravitational wave background. These constraints may eventually push string tension limits well below $G\mu \sim 10^{{-7}}$, which would make the impact of cosmic strings on the CMB anisotropy very hard to observe  \cite{Hindmarsh:2011qj,vanHaasteren:2011ni, Kuroyanagi:2012jf,Sanidas:2012ee,Sanidas:2012tf}. 

Fortunately, there seems to be a  way to overcome this potential problem (in addition to various issues with interpretation of pulsar timing results discussed in \cite{Hindmarsh:2011qj,Kuroyanagi:2012jf,Sanidas:2012ee,Sanidas:2012tf}). Indeed, it was noticed long ago that the cosmological phase transitions with string production may take place not only {\it after inflation} but also {\it during inflation} \cite{Shafi:1984tt,Kofman:1986wm,Vishniac:1986sk,Yokoyama:1989pa}. If inflation continues less than 60 e-foldings after the phase transition with string formation, it does not affect the contribution of strings on perturbations of metric on very large scale responsible for large angle CMB anisotropy. Meanwhile the number density of small string loops, giving the dominant contribution to the gravitational wave background affecting the pulsar timing, may be very strongly suppressed. This may allow one to avoid the pulsar timing limits on string tension \cite{Yokoyama:1989pa}. The same effect can help to avoid excessive particle production by oscillating cosmic string loops.

This idea was recently implemented in a particular model of chaotic inflation in supergravity \cite{Kamada:2012ag}. The authors studied F-term strings, which required introduction of 3 new scalar fields interacting with each other, in addition to the fields present in the inflationary model. 

In our paper, we will discuss a similar but more general class of models of inflation in supergravity, which allow to easily tune basic cosmological parameters such as $n_{s}$, $r$, and $f_{{NL}}$. We will see that in this class of models one can describe Abelian Higgs D-term string production during inflation by adding to the basic inflationary model just one extra scalar field with vanishing superpotential.

\section{Chaotic inflation in supergravity}

The model to be discussed here belongs to the broad class of models of chaotic inflation in supergravity developed in \cite{Kallosh:2010ug,Kallosh:2010xz,Kallosh:2011qk}, which generalized
the simplest model of this type proposed in \cite{Kawasaki:2000yn}; see also \cite{Yamaguchi:2000vm,Yamaguchi:2001pw,Kawasaki:2001as,Yamaguchi:2003fp,Brax:2005jv,Kallosh:2007ig,Kadota:2007nc,Davis:2008fv,Einhorn:2009bh,Ferrara:2010yw,Lee:2010hj,Ferrara:2010in,Takahashi:2010ky,Nakayama:2010kt,Silverstein:2008sg,McAllister:2008hb}
for a partial list of related publications.

 The class of models developed in \cite{Kallosh:2010ug,Kallosh:2010xz,Kallosh:2011qk} describes two scalar fields, $S$ and $\Phi$, with the superpotential 
 \be
{W}= Sf(\Phi) \ ,
\label{cond}
\ee
where $f(\Phi)$ can be any real holomorphic function such that $\bar f(\bar \Phi) = f(\Phi)$. Any function which can be represented by Taylor series with real coefficients has this property. The \K \, potential can be any function of the following general form
\be\label{Kminus}
K= K((\Phi-\bar\Phi)^2,S\bar S).
\ee
In this case, the \K\, potential does not depend on $\phi = \sqrt 2\, {\rm Re}\, \Phi$. Under certain conditions on the \K\, potential, inflation occurs along the direction $S = {\rm Im}\, \Phi = 0$, and the field $\phi$ plays the role of the inflaton field with the F-term potential 
\be
V_{F} = |f(\phi/\sqrt 2)|^{2}.
\ee
Note that this expression for the inflaton potential coincides with the expression of the potential of the field $\phi$ in global SUSY with the superpotential ${W}= Sf(\Phi)$. All scalar fields have canonical kinetic terms along the inflationary trajectory $S = {\rm Im}\, \Phi = 0$. 
 
 Alternatively, one may consider models with the \K\, potential $K= K((\Phi+\bar\Phi)^2,S\bar S),$
  in which case the role of the inflaton field will be played by $\chi = \sqrt 2\, {\rm Im}\, \Phi$,  with the inflaton potential 
$V_{F} = |f(\chi/\sqrt 2)|^{2}.$

The functional freedom of choice of the inflaton potential  allows to account for {\it any} set of observational data that can be expressed in terms of  two parameters $n_{s}$, $r$. The choice of the \K\, potential controls masses of the fields orthogonal to the in\-fla\-ti\-o\-na\-ry trajectory \cite{Kallosh:2010ug,Kallosh:2010xz,Kallosh:2011qk}. If mass of some of these fields are smaller than $H$ during inflation, one may use these fields as curvaton fields   \cite{curva} for generation of non-Gaussian perturbations in this class of models \cite{Demozzi:2010aj}. Alternatively, one may produce non-Gaussian perturbations in this scenario by considering interaction with vector fields $\sim \chi F_{\mu \nu}\tilde{F}^{\mu\nu}$, see \cite{Linde:2012bt} and references therein.

\section{Cosmic string production}

The simplest way to introduce cosmic strings to these models  is to add to these models a supersymmetric version of the Abelian Higgs model describing a charged scalar field $Q$, without any superpotential associated with it, interacting with the Abelian field $A_{\mu}$ with a gauge coupling constant $g$ \cite{Dvali:2003zh}. It is not necessary to introduce two charged scalars, as in F-term or D-term inflation. 
In the global SUSY limit, the potential of the field $Q$ is given by the D-term potential 
\be V_{D} = {g^{2}\over 2} \left(\bar Q Q - \xi\right)^{2}. \ee
For the purposes of the present paper, we will treat $\xi$ as a constant, referring to the extensive recent literature for the discussion of a possible nature of the FI term, see e.g. \cite{Binetruy:2004hh,Arnold:2012yi} and references therein.

We will represent the field $Q$ as $Q = {h\over \sqrt 2} e^{{i\theta}}$, where  $h$ is canonically normalized, and the phase $\theta$, as usual,  is eliminated due to the Higgs effect. In global SUSY, D-term potential makes the field $h$ tachyonic at $h = 0$, $m^{2}_{h} = -g^{2}\xi$, which leads to spontaneous symmetry breaking due to generation of a classical scalar field $h = h_{0}$, where $h_{0} = \sqrt {2\xi}$. Spontaneous symmetry breaking results in creation of the Abelian Higgs D-term strings with tension $G\mu = {\xi/ 4}$ \cite{Dvali:2003zh}.
In what follows, it will be convenient for us to express everything alternatively either in terms of $h_{0}$, or  $\xi$, or $G\mu$. In particular the D-term contribution to the mass squared of the field $h$ near $h = 0$ can be represented as $m^{2}_{h} = -{g^{2}\over 2} h_{0}^{2}$.

In supergravity, the field $h$  acquires an additional contribution to its mass due to the F-term. This contribution depends on the choice of the K{\"a}hler\, potential and the superpotential.  As a simplest example of chaotic inflation with string production, we will consider now a theory with the superpotential 
\be
W = mS\Phi
\ee
and the  \K\, potential:
\be
\mathcal{K} =  -\frac{1}{2}(\Phi-\bar\Phi)^2 +S \bar S +Q \bar Q \ .  
\ee
Under certain conditions, the fields $S$ and $\chi = \sqrt 2\, {\rm Im}\, \Phi$ can be stabilized near the inflationary trajectory $S= \chi = 0$ \cite{Kallosh:2010ug,Kallosh:2010xz}. The strength of stabilization depends on terms like  $(S \bar S)^{2}$ which can be added to the \K\, potential. 

The potential of the fields $\phi$ and $h$ at the inflationary trajectory $S= \chi = 0$ is given by
\be
V = V_{F} + V_{D} = e^{h^{2}/2}\, {m^{2}\over 2} \phi^{2} + {g^{2}\over 4} \left(h^{2} - h_{0}^{2}\right)^{2} \ .
\ee
In the regime with $h \ll 1$ which we are going to discuss below, this expression yields
\be
V =  {m^{2}\over 2} \phi^{2}\left(1 + {h^{2}\over 2}\right)  + {g^{2}\over 4} \left(h^{2} - h_{0}^{2}\right)^{2} \ .
\ee
This means that the effective mass of the field $h$ near $h = 0$ at the inflationary trajectory is given by
\be \label{higgsmass}
m^{2}_{h} = {m^{2}\over 2}\phi^{2} -{g^{2}\over 2} h_{0}^{2}\ .
\ee
For  ${m^{2}} \phi^{2} > -{g^{2}} h_{0}^{2}$ this potential has a stable minimum at $h = 0$. The point $h = 0$ becomes unstable and spontaneous symmetry breaking accompanied by cosmic string production occurs for $\phi < \phi_{c}$, where
\be
\phi_{c}^{2} = {g^{2}h_{0}^{2}\over m^{2}}  = {2g^{2}\xi\over m^{2}} = {8g^{2}G\mu \over m^{2}}\ . 
\ee

%Let us compare the values of $V_{F}$ and $V_{D}$  at $h = 0$, $\phi> \phi_{c}$: $V_{F} = {m^{2}\over 2} \phi^{2} > {g^{2}h_{0}^{2}\over 2}$, $V_{D} = {g^{2}\over 4} h_{0}^{4}$. In the simplest case when $h_{0} \ll1$ (sub-Planckian symmetry breaking), the potential energy during inflation is dominated by the F-term, so one has 
One can easily check that during inflation in our scenario, $V_{F} \gg V_{D}$ if $m \gg g\xi$. As we will see, the condition $m \gg g\xi$ is satisfied in the realistic version of our model. In this case, the Hubble constant during inflation is given by
\be\label{hubble}
H^{2} = {m^{2}\phi^{2}\over 6} \ , 
\ee
just as in the simplest chaotic inflation scenario with a quadratic potential without any D-term contribution \cite{Linde:1983gd,Linde:2005ht}. In this scenario, the number $N_{c}$ of e-folds of inflation during the process of rolling of the field down from $\phi_{c}$ is given by $N = \phi^{2}_{c}/4$. This means that if the phase transition with string formation occurs during inflation, the universe expands by the factor of 
\be
e^{N} = e^{2g^{2}G\mu/m^{2}}
\ee
since the moment of the phase transition. In the realistic version of this scenario, COBE normalization requires $m \sim 6 \times 10^{-6}$, so one has  
\be
e^{N} = e^{55 \times 10^{9} g^{2}G\mu} \ ,
\ee
where $G\mu = \xi/4$. For example, one would have $55$ e-folds of inflation since the moment of string production in the model with $10^{9} g^{2}G\mu = 1$. This would dilute all but very longest strings formed during inflation. This particular regime would require $g = 0.1$ for $G\mu = 10^{-7}$, or $g = 0.05$ for $G\mu = 4\times 10^{-7}$, which would lead to a small observable contribution of strings to the CMB anisotropy. The consistency condition $m \gg g\xi$ required for the regime  $V_{F} \gg V_{D}$ is satisfied for these values of parameters. 

Following \cite{Kamada:2012ag}, one may argue that if the phase transitions with string production occurs at a sufficiently large $\phi_{c}$, like in the example given above, the stringent pulsar constraints on the string tension may no longer apply. However, a more detailed investigation of this question is warranted since observational consequences of string formation during inflation are less explored than the consequences of the conventional mechanism of string formation after inflation.

In particular, in our discussion we made an implicit assumption that the phase transition occurs at $\phi = \phi_{c}$. Meanwhile the actual situation is more complicated. The process of string formation first begins at $\phi > \phi_{c}$ when the mass of the field $h$ becomes smaller than the Hubble constant ${m^{2}\phi^{2}\over 6}$. Later on, the tachyonic instability becomes very fast  approximately at the time when the absolute value of the tachyonic mass of the field $h$ becomes equal to H. One can easily check that in our model the first of these two moments occurs at $\phi^{2} = {3\over 2}  \phi_{c}^{2}$, and the second one occurs at $\phi^{2} \sim {3\over 4}  \phi_{c}^{2}$. 

This means, for example, that if the universe expands $e^{\phi_{c}^{2}/4} = e^{60}$ times after the field passes through the critical point, then the process of string formation begins due to generation of quantum fluctuations of the field $h$ approximately 30 e-folding {\it earlier}, and it ends approximately 15 e-folding {\it later} than one could naively expect. This also means that a tiny fraction of cosmic strings may be produced at slightly later times as well, because of rare fluctuations occasionally bringing the field $h$ back to $h = 0$. Thus we deal with a rather smooth and broad transition process, so in general one may even take the critical point $\phi_{c}$ to be well above its value corresponding to 60 e-foldings.

For a more detailed analysis of this process one may use estimates of the type made in \cite{Kamada:2012ag}. The best way to explore this process in detail is to use lattice simulations,  as it was done in our study of symmetry breaking process in hybrid inflation \cite{Felder:2000hj,Felder:2001kt}. We hope to return to a discussion of this  issue in a separate publication.

Our model of string production allows various generalizations. In particular, instead of studying models with the simplest superpotential $W = mS\Phi$, one may study a more general case with ${W}= Sf(\Phi)$. After some algebra, one finds  that for any holomorphic function $f(\Phi)$ the mass of the field $h$ during inflation is given by
\be
m^{2}_{h} = V_{F}(\phi)  -{g^{2}\over 2} h_{0}^{2}\ ,
\ee
where $V_{F}(\phi) = |f(\phi/\sqrt 2)|^{2}$. This generalizes our previous result (\ref{higgsmass}). The critical point of the phase transition phase transition with string production corresponds to the moment when $V_{F}(\phi)$ becomes smaller than ${g^{2}\over 2} h_{0}^{2}$. Whereas the description of inflation does depend on the choice of the function $f(\Phi)$, the main qualitative results concerning string production during inflation remain valid for a broad choice of inflationary potentials $V_{F}(\phi)$. 

Further generalizations of this model are possible. For example, by adding extra terms like $(S\bar S)^{2}$ to the  \K\, potential, one can control the mass of the field $S$ \cite{Kallosh:2010ug,Kallosh:2010xz}, which may play the role of the curvaton field in this scenario  \cite{Demozzi:2010aj}. Meanwhile adding an extra term $\kappa S\bar S Q\bar Q$ to the \K\, potential alters our expression for the Higgs mass, which becomes $m^{2}_{h} = (1-\kappa)\, V_{F}(\phi)  -{g^{2}\over 2} h_{0}^{2}$ \cite{Linde:2012bt}. This modification controls the position of the critical point of the phase transition with string formation, and therefore the duration of inflation after it.
As we already mentioned, one can obtain an additional source of non-Gaussianity by considering the \K\, potential $K= K((\Phi+\bar\Phi)^2,S\bar S),$ and adding interaction with vector fields $\sim \chi F_{\mu \nu}\tilde{F}^{\mu\nu}$  \cite{Linde:2012bt}. 

To conclude, in the context of the general class of chaotic inflation models discussed above one can consistently fit {\it any} set of observational parameters $n_{s}$ and $r$, introduce a controllable amount of non-Gaussianity, and also add a small but potentially observable contribution of cosmic strings to the CMB anisotropy.

\subsection*{Acknowledgments}
I am very grateful to R. Kallosh, S. Mooij and E. Pajer for many enlightening discussions. This work was
 supported by NSF grant PHY-0756174. 

%%%%%%%%%%%%%%%%%%%%%%%%%%%%%%%%%%%%%%%%%%%%%%%%%%

\end{document}